\documentclass[jkps,twoside,twocolumn,showpacs,showkeys]{revtex4}
\usepackage{graphicx}
\usepackage{amssymb}
\usepackage{amsmath}
\usepackage{bm}

\newcommand{\ttbb}{t\bar{t}b\bar{b}}

\newcommand{\ttjj}{t\bar{t}jj}
\newcommand{\ttH}{t\bar{t}H}
\newcommand{\ttbar}{t\bar{t}}
\newcommand{\bbbar}{b\bar{b}}

\newcommand{\sttbb}{\sigma_{t\bar{t}b\bar{b}}}
\newcommand{\sttjj}{\sigma_{t\bar{t}jj}}

\newcommand{\GeVcc}{{\rm GeV}/c^2}
\newcommand{\GeVc}{{\rm GeV}/c}
\newcommand{\pt}{p_{\rm T}}

\begin{document}
\setcounter{page}{0}
\title[]{Study of the Top-quark Pair Production 
in Association with a Bottom-quark Pair from Fast Simulations at the LHC}
\author{Young Kwon \surname{Jo}}
\affiliation{Department of Physics, Korea University, Seoul
135-701}
\author{Su Yong \surname{Choi}}
\affiliation{Department of Physics, Korea University, Seoul
135-701}
\author{Tae Jeong \surname{Kim}}
\email{tae.jeong.kim@jbnu.ac.kr}
\affiliation{Department of Physics, Chonbuk National University, Jeonju
561-756}
\author{Youn Jung \surname{Roh}}
\affiliation{Department of Physics, Korea University, Seoul
135-701}

\date[]{16 June 2015 }

\begin{abstract}
A large number of top quarks will be produced at the Large Hadron Collider (LHC) during the Run II period. 
This will allow us to measure the rare processes from the top sector in great details.
We present a study of top-quark pair production in association with a bottom-quark pair ($\ttbb$) 
from fast simulations for the Compact Muon Solenoid (CMS) experiment. 
The differential distributions of $\ttbb$ are compared with the top-quark pair production with two additional jets ($\ttjj$) and 
with the production in association with the Higgs ($\ttH$),
where the Higgs decays to a bottom-quark pair.
The significances of the $\ttbb$ process in the dileptonic and the semileptonic decay modes are calculated with the data corresponding to an integrated luminosity of 10 fb$^{-1}$,
which is foreseen to be collected in the early Run II period.
This study will provide an important input in searching for new physics beyond the standard model, 
as well as in searching for the $\ttH$ process where the Yukawa coupling with the top quark can be directly measured. 
\end{abstract}

\pacs{14.65.Ha}

\keywords{Top physics, Delphes}

\maketitle

\section{INTRODUCTION}

After the discovery of the Higgs boson in 2012, the phase space of searching for the long-sought Higgs boson has been
replaced by the phase space of measuring the properties of this new boson, including Yukawa coupling with the top quark. 
The fact that the top quark has the largest mass in the standard model has convinced us of its important role for checking 
the consistency of the Higgs boson with the standard model predictions. 
One of the most promising channels for a direct measurement of the top quark's Yukawa coupling
is the production of a top-quark pair in association with the Higgs boson.
The Higgs boson which decays to $\bbbar$ in the standard model will lead to a $\ttbb$ final state.
This final state, which has not yet been observed using Run I (2010-2012) data, has an irreducible
nonresonant background from the production of a top quark pair in
association with two jets faking b jets or with a b quark pair.  

The expected cross section for $\ttH$ in pp collisions at $\sqrt{s}$ = 8 TeV to next-to-leading order (NLO)
is $0.128^{+0.005}_{-0.012}$ (scale) $\pm$ 0.010 pb (PDF+$\alpha_\mathrm{S}$)~\cite{yellowreport}.
Calculations of $\sttjj$ and $\sttbb$ have been performed to NLO precision.
$\sttjj$ and $\sttbb$ predictions at $\sqrt{s}$ = 8 TeV are
\mbox{$\sttjj = 21.0 \pm 2.9$ (scale) pb} and \mbox{$\sttbb = 0.23 \pm 0.05$ (scale) pb}~\cite{ttbbQCD}.
The ratio of $\sttbb$ to $\sttjj$ is $0.011 \pm 0.003$.  
In this calculation, the additional jets must  
have transverse momenta \mbox{$\pt$ $>$ 40 $\GeVc$} and absolute pseudorapidity $|\eta|$ $<$ 2.5.
The dominant uncertainties in these calculations are from the factorization and the renormalization scales
caused by the presence of two very different scales in this process, the top quark mass and the jet $\pt$ scales.

The first measurements of the cross sections $\sttbb$ and $\sttjj$ and their ratios were presented by the Compact Muon Solenoid (CMS) experiment 
at the CERN Large Hadron Collider (LHC)
by using the full data sample of pp collisions at a center-of-mass 
energy of 8 TeV, which corresponds to an integrated luminosity of 20 fb$^{-1}$. The measured cross sections of $\sttbb$ and $\sttjj$ in the same phase space as in the theory calculation are
\mbox{$\sttbb$ = 0.36 $\pm$ 0.08 (stat.) $\pm$ 0.10 (syst.) pb} and
\mbox{$\sttjj$ = 16.1 $\pm$ 0.7 (stat.) $\pm$ 2.1 (syst.) pb}, respectively~\cite{ttbbCMS}.
The measured cross section ratio of $\sttbb$ to $\sttjj$ is 0.022 $\pm$ 0.004 (stat.) $\pm$ 0.005 (syst.),
which is compatible within 1.6 standard deviation with the theory prediction of $0.011 \pm 0.003$.

A large number of top quark candidates is expected to be produced at the LHC 
even during the early Run II period (2015-2017).
This will allow us to measure these processes in great detail.
Therefore performing the same measurement at 13 TeV is important because
the experimental measurements of $\sttjj$ and $\sttbb$ production can provide a good test of NLO QCD theory
and an important input about the main background in the search for the $\ttH$ process.
This study will also provide useful information about the main background in the search for new physics beyond the standard model, 
such as the flavor-changing neutral current process where one of the top quarks decays
to an up (or charm) quark and a Higgs boson and the Higgs boson decays to a bottom-quark pair.

In this paper, we present a fast-simulation study of the $\ttbb$ process and compare it with the $\ttjj$ and the $\ttH$ processes.
The differential distributions of additional b jets from $\ttbb$ events are compared with the differential distributions of b jets from  $\ttjj$ events and from $\ttH$ events
where the Higgs decays to a bottom-quark pair.

\section{SAMPLES}

The simulated pp collision data samples for three processes, $\ttbb$, $\ttjj$ and $\ttH$, are produced
separately at a center-of-mass energy of 13 TeV.  
The $\ttbb$ and $\ttH$ samples  
are generated by using the MadGraph5\_aMC@NLO (v2.1.2)~\cite{aMCNLO} framework at the NLO level 
and are further interfaced with PYTHIA (v8.185) for the hadronization.
The $\ttjj$ samples for the differential distributions 
are generated by using MadGraph5 at leading order due to the limit of computer resources 
interfaced with PYTHIA (v6.428) for the hadronization.    
For $\ttbb$ and $\ttH$ samples, 100K events are produced 
while for $\ttjj$ sample, 1M events are produced.    

A transverse momentum threshold of 20 $\GeVc$ and a pseudo-rapidity $|\eta|$ $<$ 2.5 for the additional jets 
that are not from top quarks are applied at the production level.
The additional jets in the $\ttjj$ events include b quarks as well as light flavor quarks and a gluon.
For the $\ttbb$ events, the events are generated in a 4-flavor scheme, where the b quark is treated as having the mass.
In the $\ttbb$ events, there must be at least two additional b jets with a threshold of 20 $\GeVc$.
In the $\ttH$ events, the decay of the Higgs boson is handled in PYTHIA. 
The cross sections of the $\ttbb$, $\ttjj$ and $\ttH$ processes at $\sqrt{s} $ = 13 TeV are calculated to NLO within the MadGraph5\_aMC@NLO framework. 
The cross sections are 224 $\pm$ 1.8 pb for the $\ttjj$ process, 4.7 pb for the $\ttbb$ process, and 0.32 pb for the $\ttH$ process at $\sqrt{s}$ = 13 TeV. 
These cross sections are used to calculate the significance by normalizing the number of events to data corresponding to 
an integrated luminosity of 10 fb$^{-1}$. 

The generated events are processed for the detector simulation with the DELPHES package (v3.1.2)~\cite{delphes} for the CMS detector.
Similar to the CMS reconstruction, the objects from the particle-flow algorithm implemented in DELPHES are used throughout 
this analysis. 
The pileup events are available and can be merged in the simulated events in the DELPHES package. 
However, in this analysis, we assume that pileup mitigation, which will be developed
at the CMS experiment, can reduce the effect from pileup events significantly. 
It is also important to understand the physics difference in the case of no pileup events.
Therefore, we focus on only the physics under the condition that there is no pileup effect. 

In the DELPHES fast simulation, the final momenta of all the physics objects, 
such as electrons, muons and jets, are smeared as a function of $\pt$ and $\eta$ so that 
they can represent the detector effects in the CMS experiment. 
The reconstruction efficiencies of the electrons, muons and jets are also parameterized as functions of $\pt$ and $\eta$
based on information from the measurements using Run 1 data. 
The muon identification efficiency is set to 95\% for the muons with 
momenta $\pt$ $>$ 10 $\GeVc$ and $\pt$ $<$ 100 $\GeVc$.
The electron identification efficiency is set to 95\% for $|\eta|$ $>$ 1.5 and 85\% for 1.5 $<$ $|\eta|$ $<$ 2.5     
The isolated muons and electrons are selected by applying a relative isolation of $I_{rel}$ $<$ 0.1, where $I_{rel}$ 
is defined as the sum of the surrounding energy from the particle-flow tracks, photons and neutral hadrons 
divided by the transverse momentum of the muon or electron.  
The particle-flow jets used in this analysis are clustered by using the particle-flow tracks and particle-flow towers.
If the jet is already reconstructed as an isolated electron, muon or photon, the jet is excluded from further consideration.  
The b-tagging efficiency parameterized as a function of $\pt$ and $\eta$ of the jet
ranges from 20\% to 50\%.
The fake b-tagging rate from the light flavor jet is set to 0.1\%, 
which corresponds to the tight-working point in the CMS measurement~\cite{cmsbjet}. 

\section{DILEPTON ANALYSIS}

In order to constrain the phase space to that we can experimentally measure in the dileptonic decay mode, 
we applied the following event selections at the reconstruction level.  
Events should have at least two leptons with $\pt > 20$ $\GeVc$ and $|\eta|$ $<$ 2.4 (S1).
Four reconstructed jets with $\pt > 30$ $\GeVc$ and $|\eta|$ $<$ 2.5 are required (S2).
Two b-tagged jets are required to select the $\ttbar$ events (S3). 
After this selection step, based on the experimental result from the CMS experiment~\cite{dilepton}, we would still have remaining backgrounds from 
single-top events and Drell-Yan events.
In order to remove these possible remaining backgrounds and to be sure we have only $\ttjj$, $\ttbb$ and $\ttH$ events
after the final selection, 
we further require the event to have one more b-tagged jet, adding up to at least three b-tagged jets (S4).   

In the best scenario where we can identify the origin of the b jets, we can see the potential difference in the kinematic distributions of the additional jets 
not from top quarks for the $\ttbb$, $\ttjj$ and $\ttH$ processes.
For this purpose, we rely on the Monte Carlo information to identify the origin of the jets as to whether or not they come from a top quark.  
The additional jets not from top quarks are identified by using the geometric information
$\Delta R (j,q) = \sqrt{ \phi^2 + \eta^2 }$, requiring $\Delta R$ $<$ 0.5, where $j$ denotes jets at the reconstruction level and 
$q$ denotes a quark that does not originate from top quarks.   
If the jet matches the b quark, it is treated as a b jet. Otherwise, the jet is treated as a light flavor jets. 

\begin{figure}
\includegraphics[width=6cm]{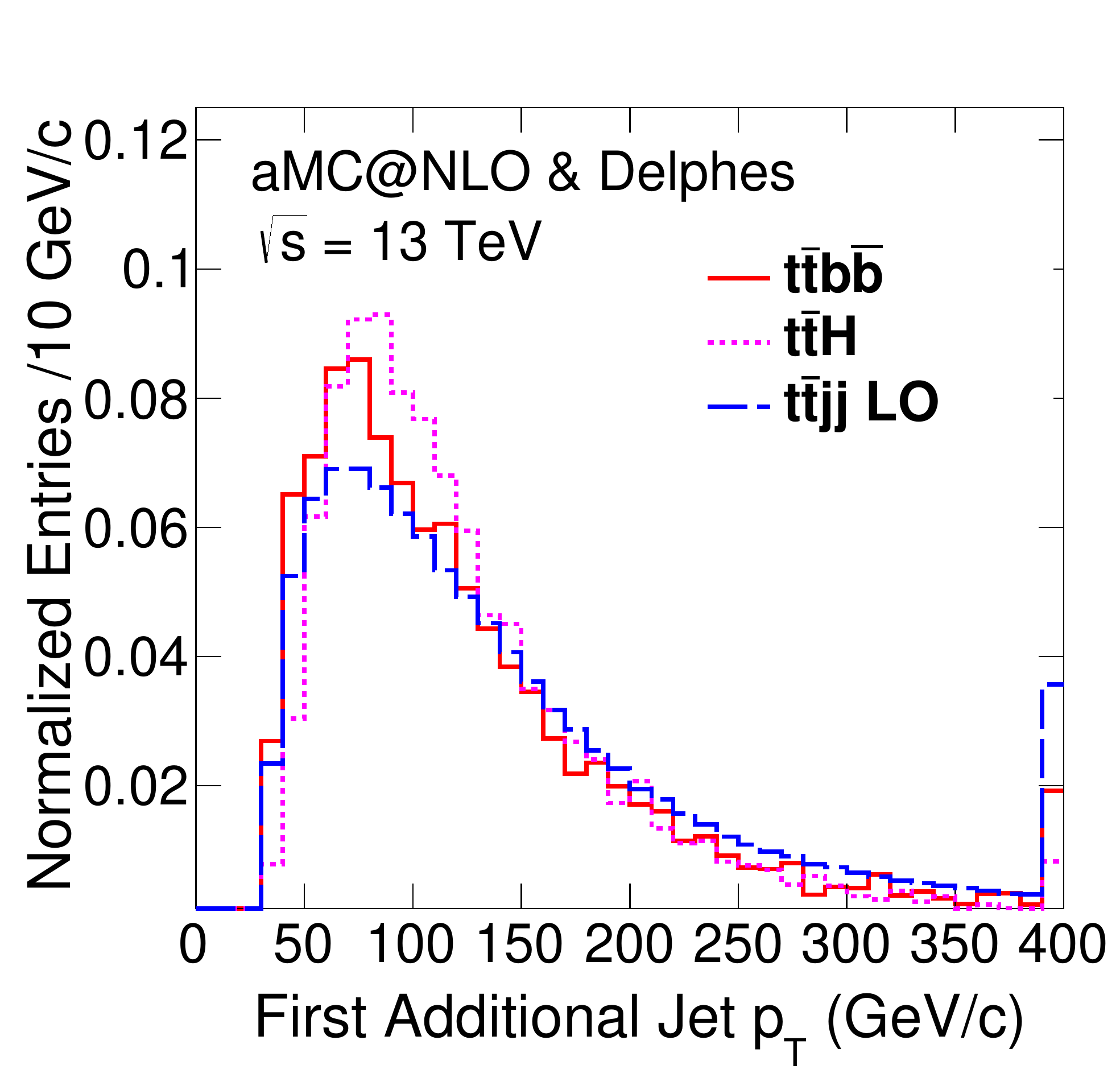}
\includegraphics[width=6cm]{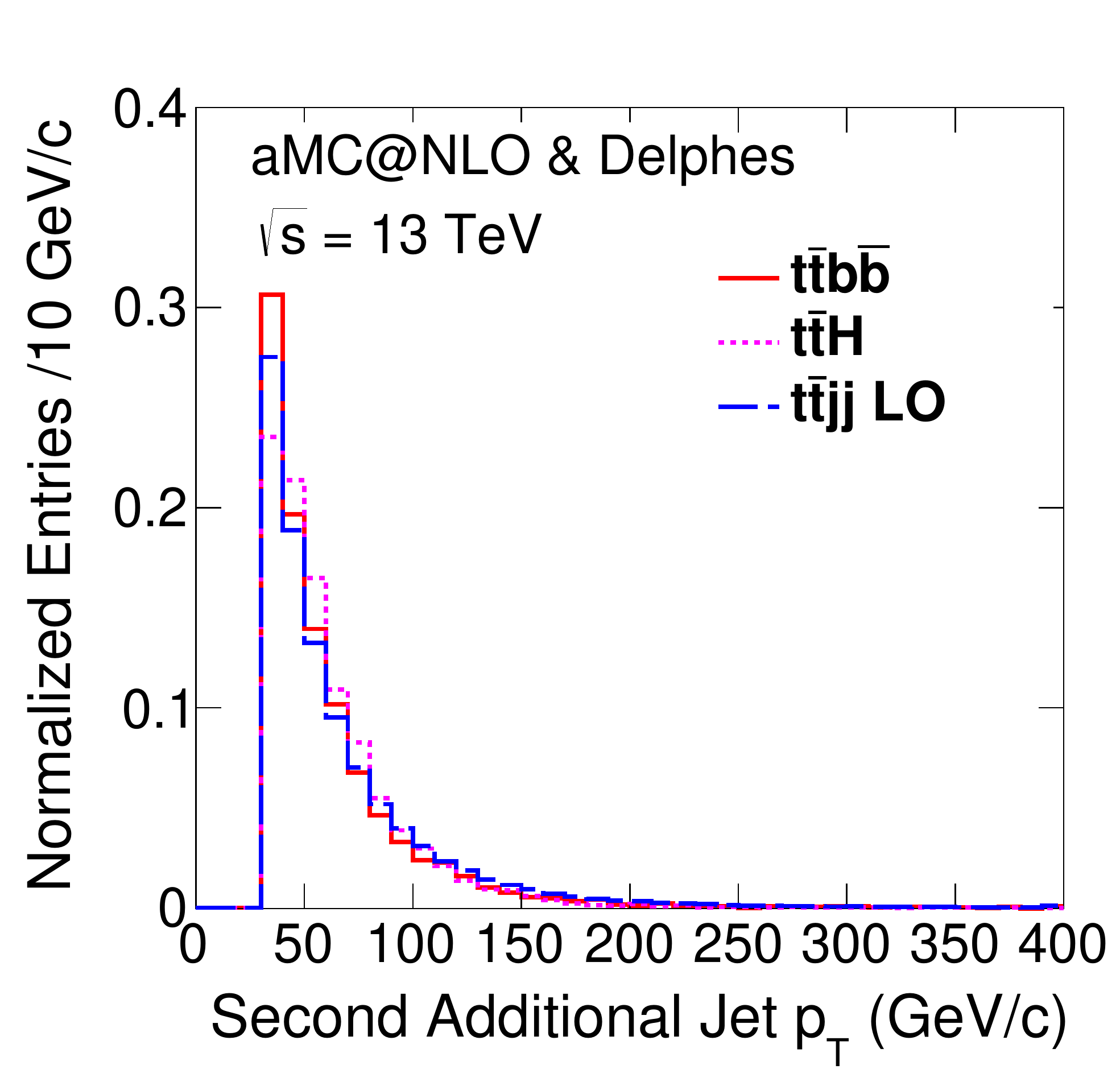}\\
\includegraphics[width=6cm]{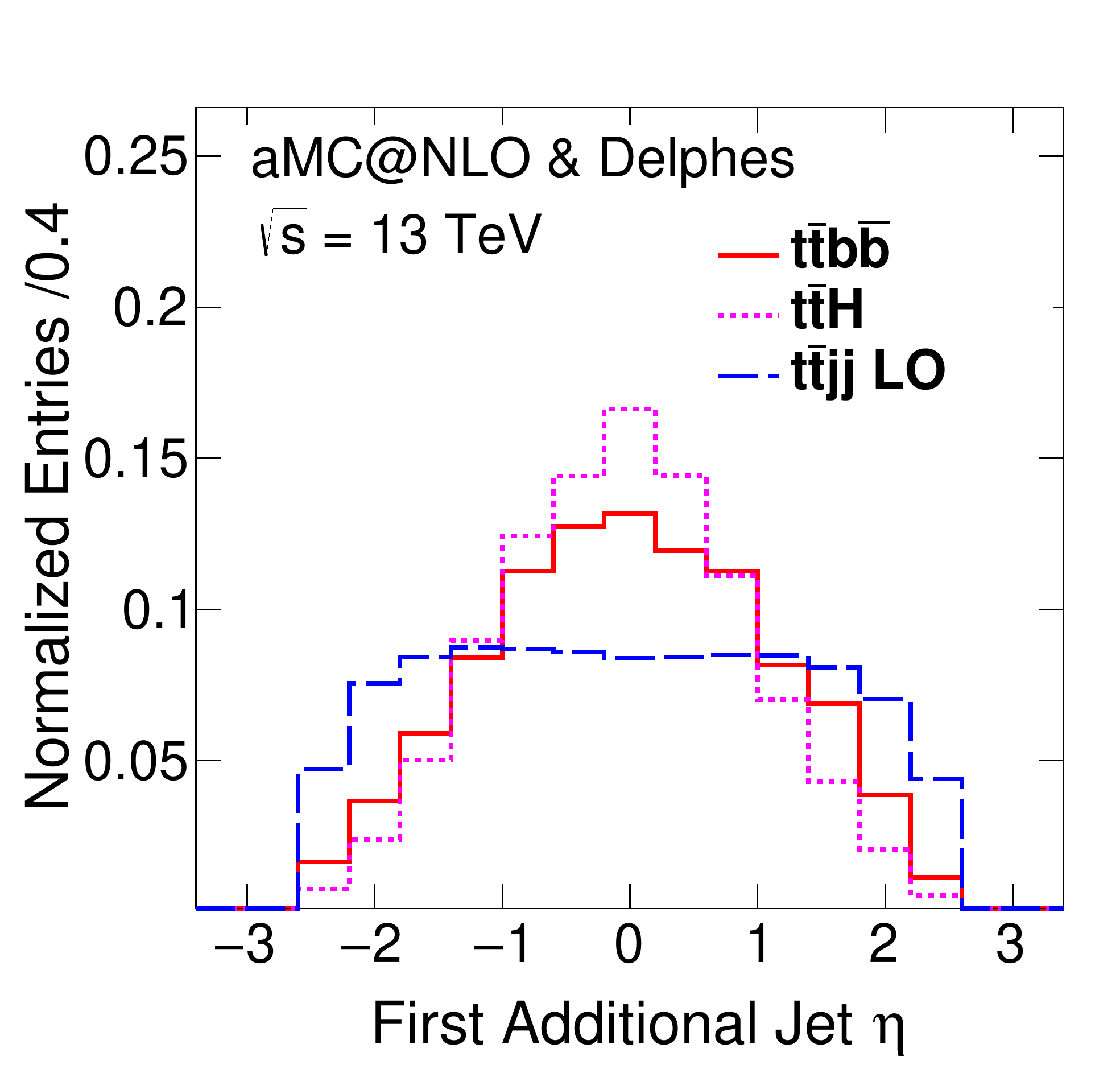}
\includegraphics[width=6cm]{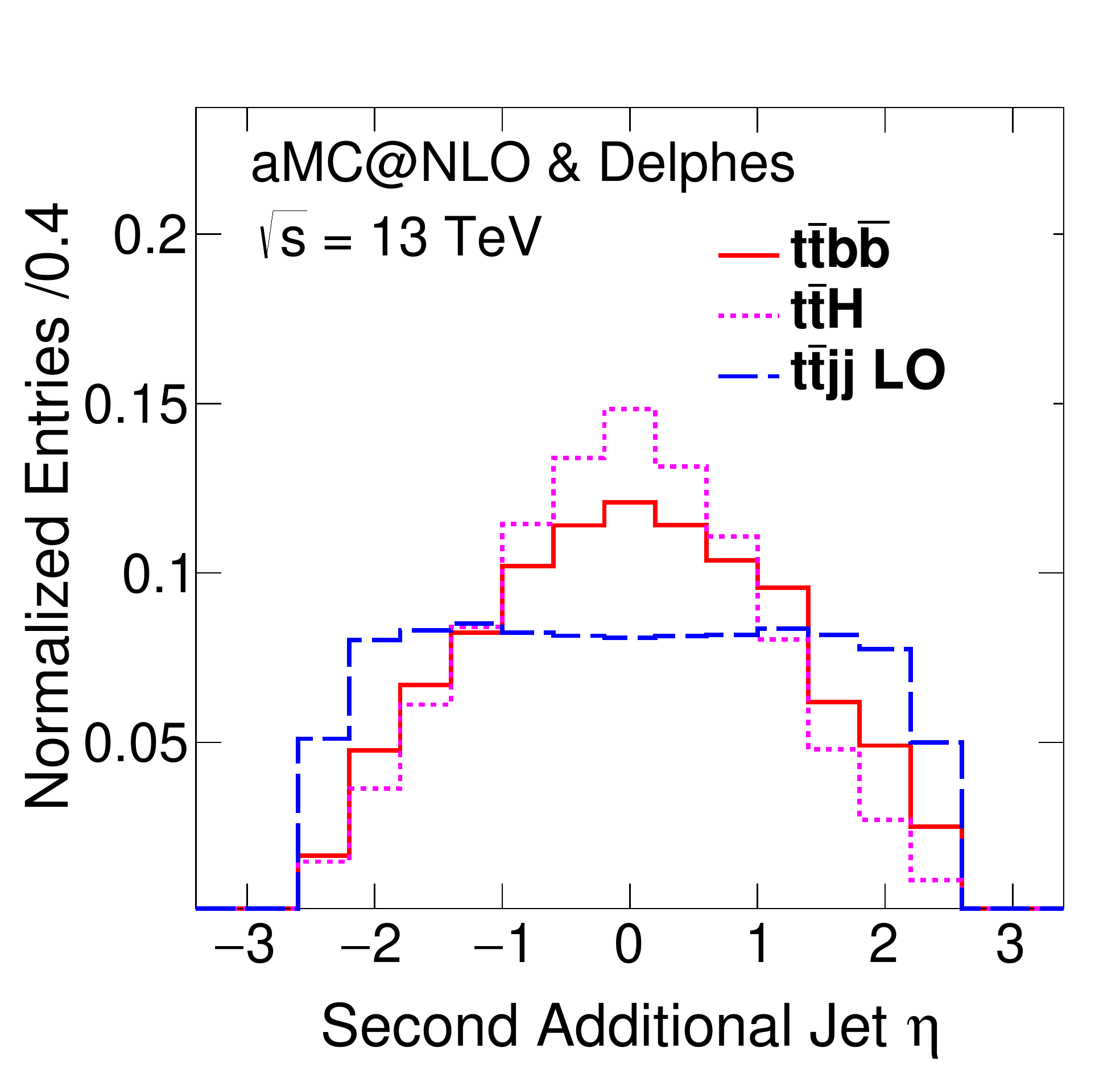}\\
\caption{Jet $\pt$ (top) and $\eta$ distributions (bottom) for the first and the second additional jets which are not from top quarks are shown
in the dileptonic decay mode.
The red solid line indicates the $\ttbb$ process. The purple dotted line shows the $\ttH$ process. 
The $\ttjj$ process is indicated by a blue dashed line.  
}\label{fig1_1}
\end{figure}

\begin{figure}
\includegraphics[width=6cm]{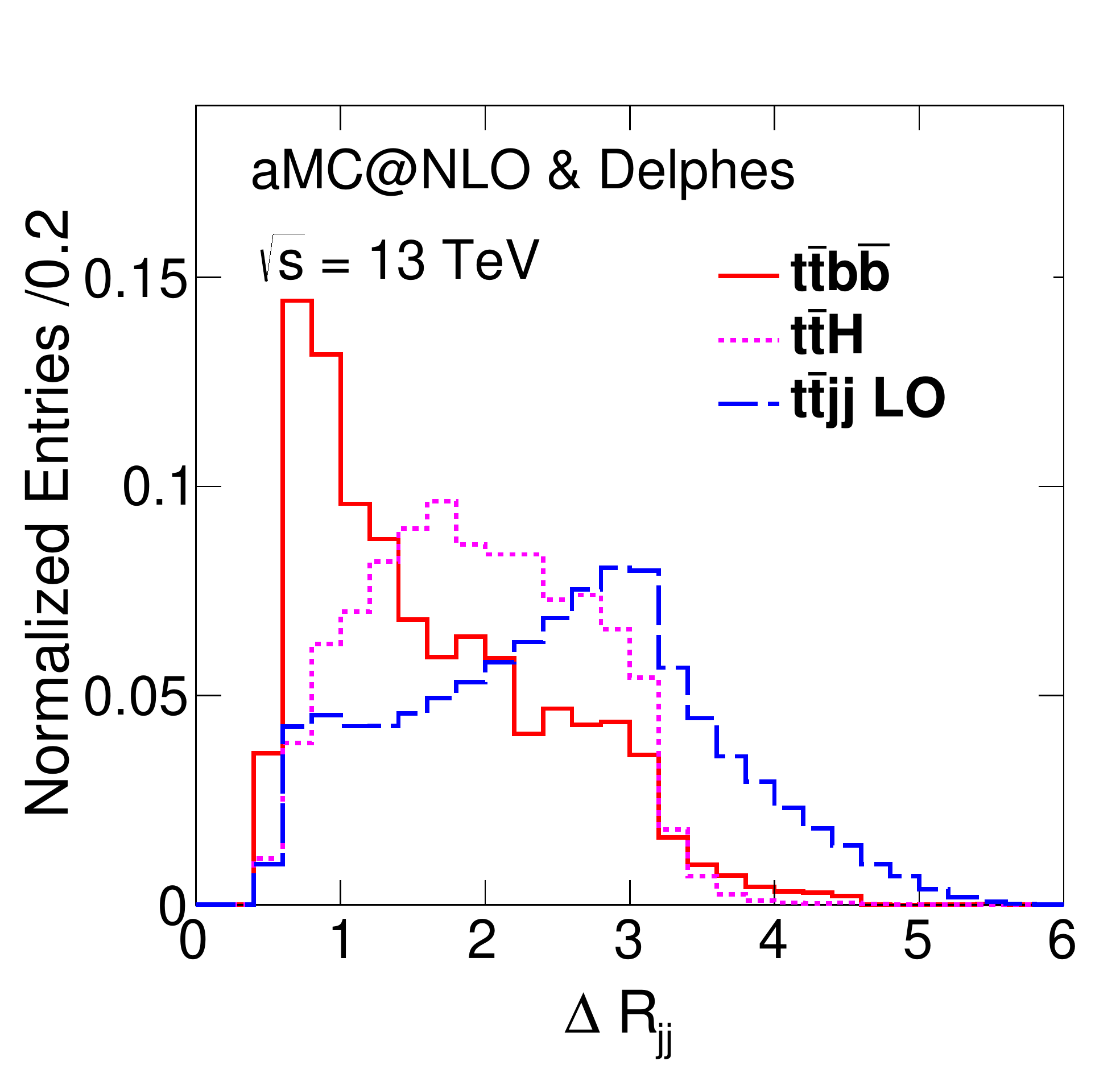}
\includegraphics[width=6cm]{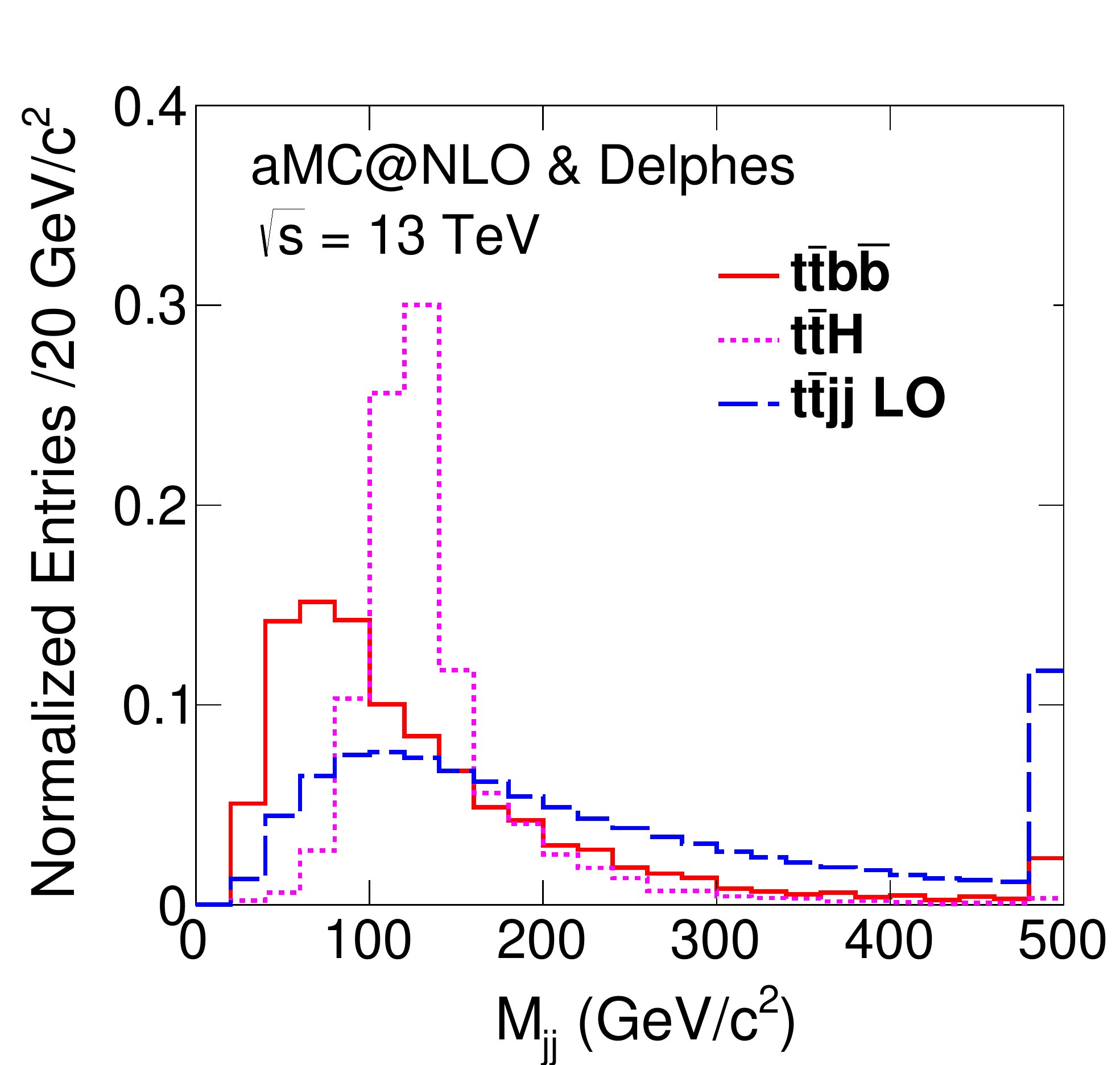}
\caption{
Invariant mass (left) and $\Delta R$ (right) distributions for two additional jets not from top quarks for the $\ttbb$, $\ttjj$ and $ttH$ processes are shown
in the dileptonic decay mode.
The red solid line indicates the $\ttbb$ process. The purple dotted line shows the $\ttH$ process.
The $\ttjj$ process is indicated by a blue dashed line.
}\label{fig1_2}
\end{figure}

Plots for this analysis are shown at the preselection owing to the lack of statistics in our simulated samples and to the tight selection based 
on the assumption that the distributions would not be significantly different after the event selections.
Figure~\ref{fig1_1} shows the transverse momentum and pseudo-rapidity distributions for two jets that are not from top quarks.
The graphs in Fig.~\ref{fig1_1} show that the additional b jets in the $\ttbb$ event 
tend to have softer $\pt$ compared to those from the $\ttH$ and the $\ttjj$ events.
The pseudo-rapidity distributions from the $\ttbb$ events are shown to be more central than those from the other processes. 
The invariant mass and $\Delta R$ distributions for these two jets are also presented in Fig.~\ref{fig1_2}. 
As expected, the invariant mass of the two b jets from the Higgs boson in the $\ttH$ process has a clear peak around 125 $\GeVcc$, 
the simulated mass in the $\ttH$ sample, while that is not the case for the other two processes, $\ttbb$ and $\ttjj$. 
The $\Delta R$ distribution shows a clear distinguishing feature. The additional jets from the $\ttbb$ process 
have a narrow angle between jets while those from the $\ttjj$ events have a wider angle and those from the $\ttH$ events are in the middle. 

\begin{table}
\caption{
Expected number of events in the dileptonic decay mode normalized to the data corresponding to an integrated luminosity of 10 fb$^{-1}$
with the NLO cross sections. The renormalization and the factorization scale uncertainties are shown only at the final selection.
The acceptance ($\epsilon$) after the final selection is also presented. 
}
\begin{ruledtabular}
\begin{tabular}{cccccc}
Process & two leptons & jets $\geq$ 4 & b-tag $\geq$ 2 & b-tag $\geq$ 3 & $\epsilon$ (\%) \\
\colrule
 $\ttbb$ & 520 & 308 & 118  & 33 $\pm$ 0.8 &  1.5 \\
 $\ttjj$ & 30183 & 10285 & 1712  & 97 $\pm$ 16  & 0.09  \\
 $\ttH$ & 39 & 21 & 7.8  & 2.0 $\pm$ 0.1  & 1.3\\
\end{tabular}
\end{ruledtabular}
\label{table1}
\end{table}

Table~\ref{table1} shows the expected number of events at each event selection step normalized to
the data corresponding to an integrated luminosity of 10 fb$^{-1}$ by using the NLO cross sections for each process.
The acceptance after the final selection is also presented. 
The expected number of events from the $\ttbb$ process after full selection is 33 events. 
The expected contributions from the $\ttH$ process and the $\ttjj$ process after the full selection are 2.0 events and 97 events, respectively.
The significance of $s/\sqrt{s+b}$, where s is the number of signal events ($\ttbb$) and b is the sum of $\ttjj$ and $\ttH$ background events, 
is 2.9 without taking into account the systematic uncertainty.

\section{LEPTON + JETS ANALYSIS}

We also performed a study of the semileptonic decay mode. 
The phase space in this study is also constrained to the semileptonic decay mode in what 
we can make experimental measurement.
The following event selections at the reconstruction level are applied.
Events should have exclusively one lepton with $\pt > 40$ $\GeVc$ and $|\eta|$ $<$ 2.4 (S1).
At least six reconstructed jets with $\pt > 30$ $\GeVc$ and $|\eta|$ $<$ 2.5 are required (S2).
Two b-tagged jets are required to select the $\ttbar$ events (S3).
In order to be sure that we have $\ttjj$, $\ttbb$ and $\ttH$ events after the final event selection, 
we further require each event to have one more b-tagged jet, adding up to at least three b-tagged jets (S4).

The same strategy is applied in the semileptonic decay mode. 
We identified two jets not from top quarks by using the Monte Carlo information. 
The additional jets not from top quarks are identified by using the geometric information
$\Delta R (j,q)$ $<$ 0.5 in the same fashion as were done in the dileptonic decay mode.

\begin{figure}
\includegraphics[width=6cm]{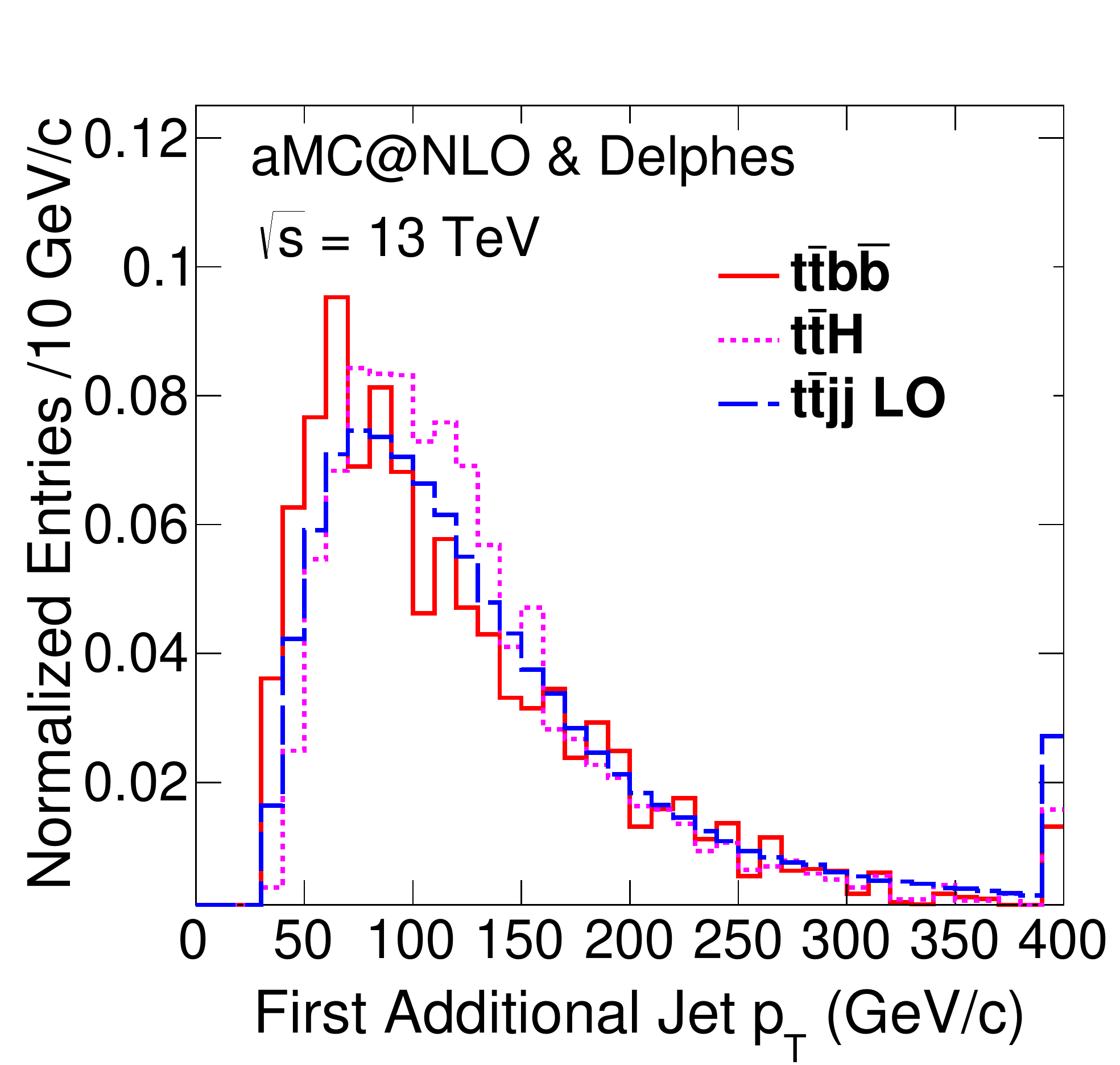}
\includegraphics[width=6cm]{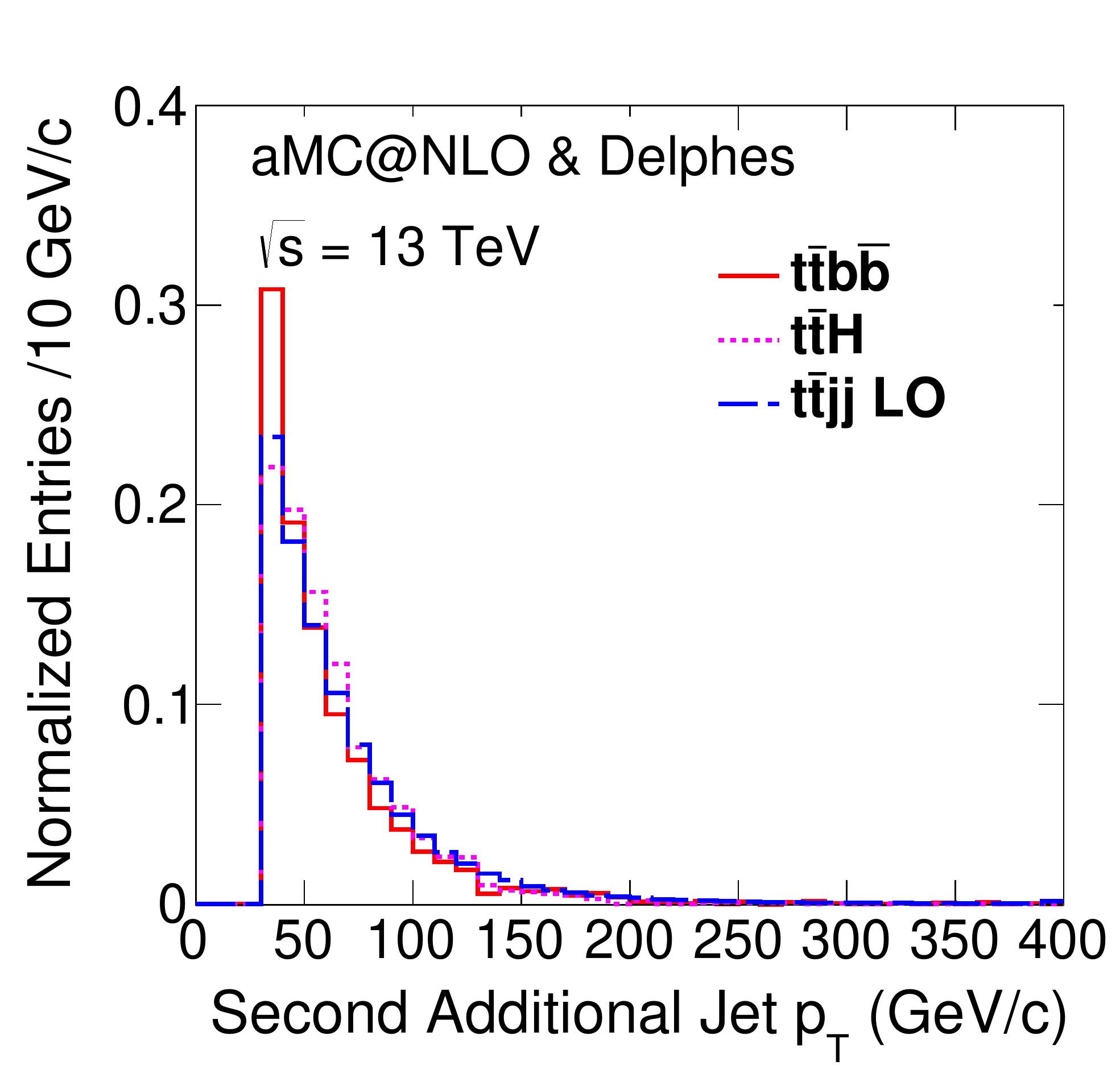}\\
\includegraphics[width=6cm]{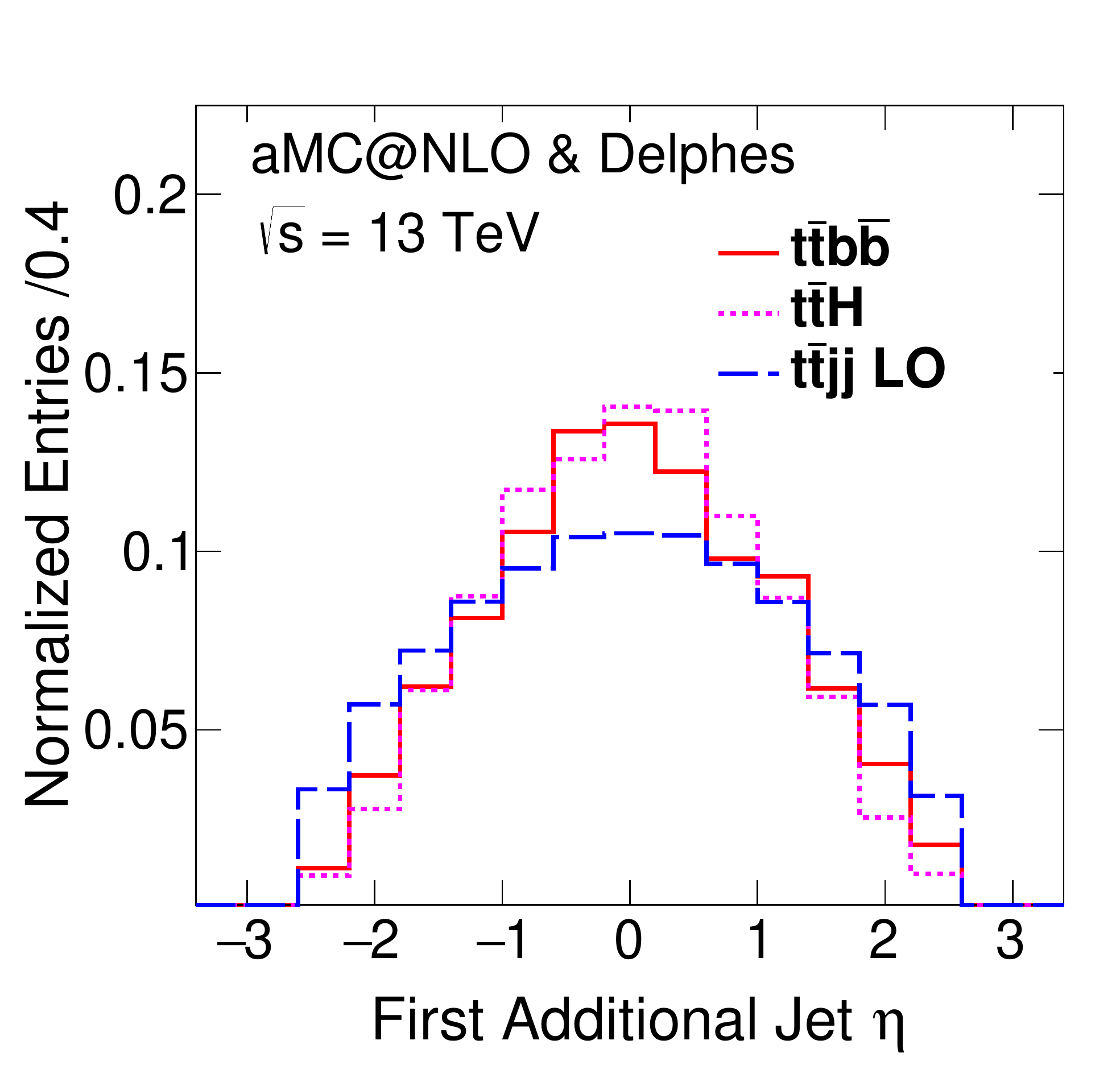}
\includegraphics[width=6cm]{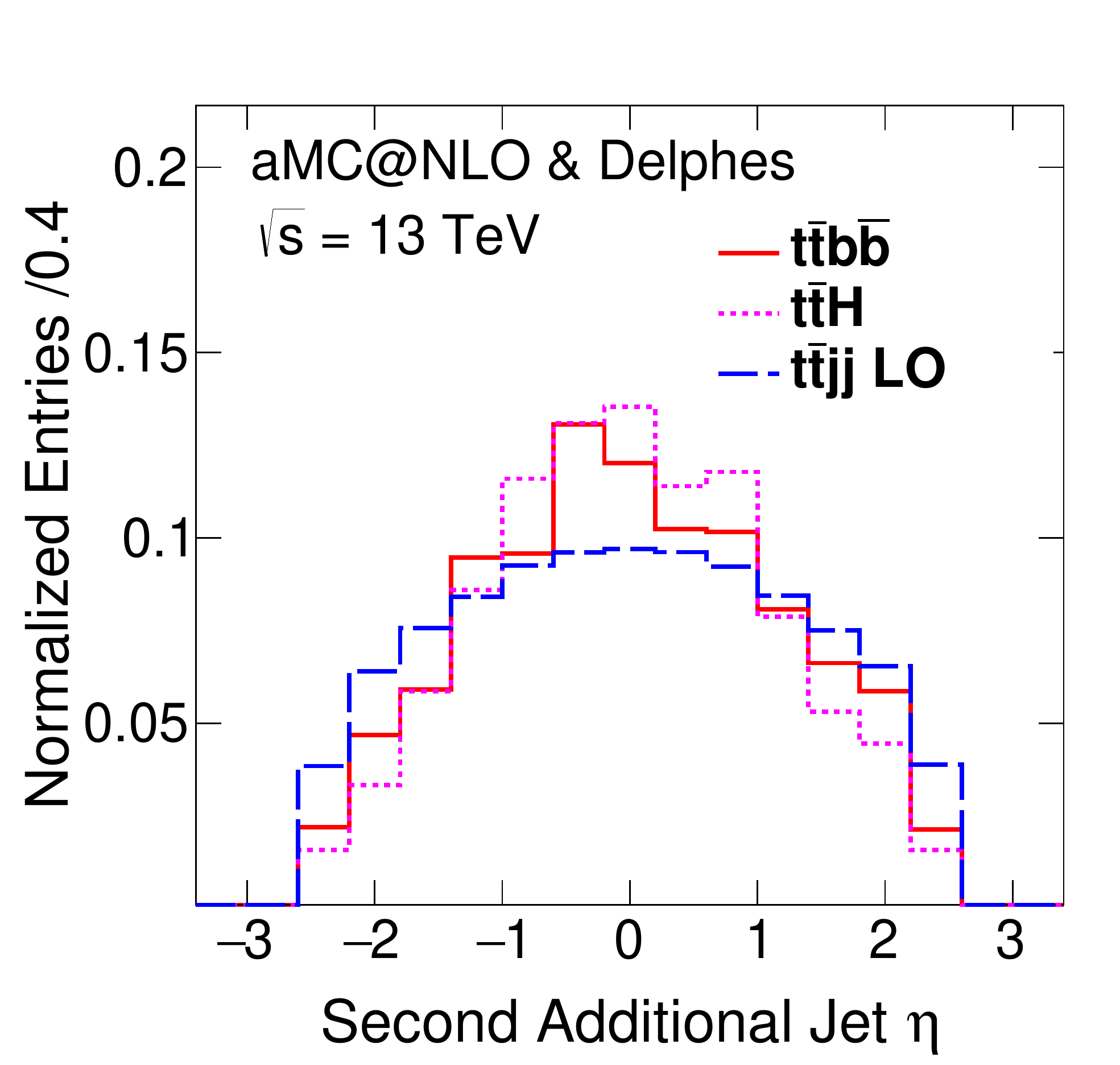}\\
\caption{
Jet $\pt$ (top) and $\eta$ distributions (bottom) for the first and the second additional jets which are not from top quarks are shown
in the semileptonic decay mode.
The red solid line indicates the $\ttbb$ process. The purple dotted line shows the $\ttH$ process. 
The $\ttjj$ process is indicated by a blue dashed line.
}\label{fig2_1}
\end{figure}

\begin{figure}
\includegraphics[width=6cm]{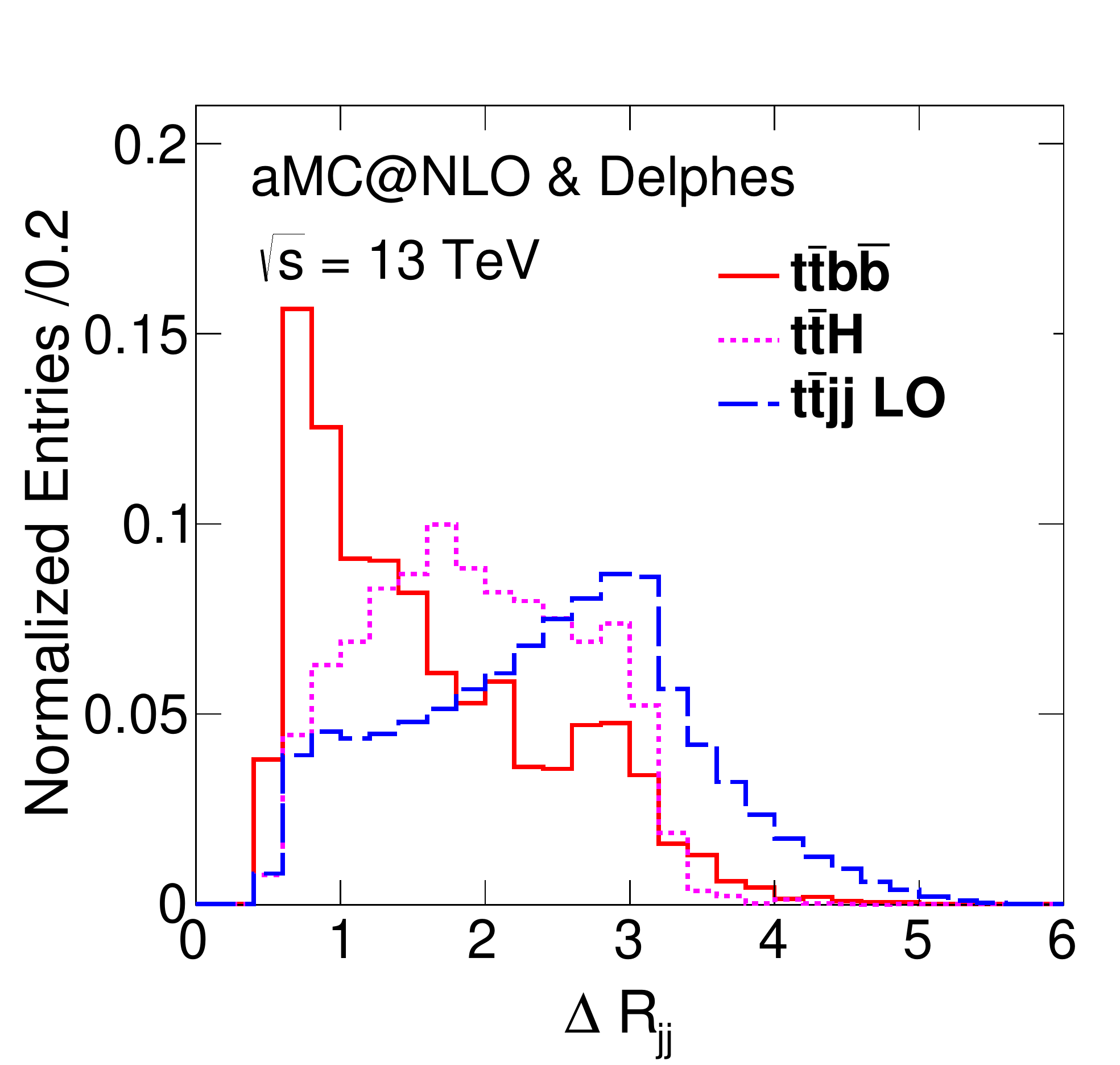}
\includegraphics[width=6cm]{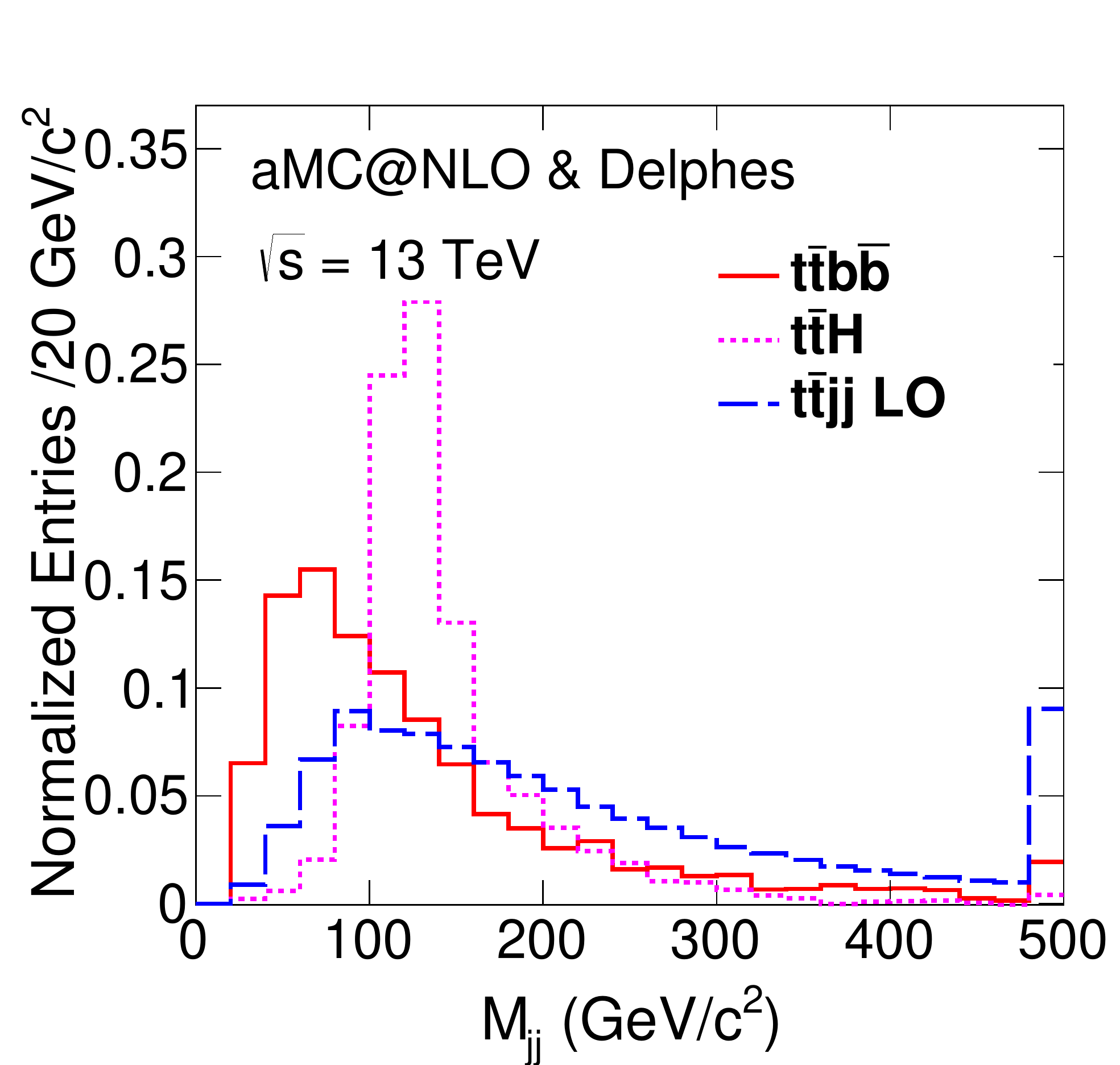}
\caption{
Invariant mass (left) and $\Delta R$ (right) distributions for two additional jets not from top quarks for the $\ttbb$, $\ttjj$ and $ttH$ processes are shown
in the semileptonic decay mode.
The red solid line indicates the $\ttbb$ process. The purple dotted line shows the $\ttH$ process.
The $\ttjj$ process is indicated by the blue dashed line.
}\label{fig2_2}
\end{figure}

Plots in this analysis are shown at the preselection owing to the lack of statistics in our simulated samples and to the tight selection based
on the assumption that the distributions would not be significantly different after the event selections.
Figures~\ref{fig2_1} show the transverse momentum and pseudo-rapidity distributions for two jets that are not from top quarks.
These show that the additional b jets in the $\ttbb$ and $\ttH$ events tend to have softer $\pt$ and central spectra compared to those from $\ttjj$ events.
The invariant mass and the $\Delta R$ distributions for these two jets are also presented in Fig.~\ref{fig2_2}. 
As expected, the invariant mass of the two b jets from the Higgs boson in the $\ttH$ process has a clear peak around 125 $\GeVcc$,
the simulated mass in the $\ttH$ sample, while that is not the case for the other two processes, $\ttbb$ and $\ttjj$.
The $\Delta R$ distribution shows a clear distinguishing feature. The additional jets from $\ttbb$
have a narrow angle between jets while for the $\ttjj$ events, they have a wider angle and for the $\ttH$, they are in the middle.

The distributions have features similar to these of the distributions in the dileptonic decay mode.
However, the pseudorapidity of two additional jets from $\ttjj$ events are shown to be more central 
compared to the one for the dileptonic decay mode.   

\begin{table}
\caption{
Expected number of events in the semileptonic decay mode normalized to the data corresponding to an integrated luminosity of 10 fb$^{-1}$
with the NLO cross sections. The renormalization and the factorization scale uncertainties are shown only at the final selection.
The acceptance ($\epsilon$) after the final selection is also presented.
}
\begin{ruledtabular}
\begin{tabular}{cccccc}
Process & one lepton & jets $\geq$ 6 & b-tag $\geq$ 2 & b-tag $\geq$ 3 & $\epsilon$ (\%)\\
\colrule
 $\ttbb$ & 5357 & 2138 & 970  & 285 $\pm$ 27 & 0.6 \\
 $\ttjj$ & 273785 & 47297 & 12238  & 2149 $\pm$ 549  & 0.04 \\
 $\ttH$ & 508 & 163 & 63 & 21 $\pm$ 2  & 0.5 \\
\end{tabular}
\end{ruledtabular}
\label{table2}
\end{table}

Table~\ref{table2} shows the expected number of events at each event selection step normalized to
the data corresponding to an integrated luminosity of 10 fb$^{-1}$ by using the NLO cross sections for each process.
The acceptance after the final selection is also presented.
The expected number of events from the $\ttbb$ process after the full selection is 285 events. 
The expected contributions from the $\ttH$ process and the $\ttjj$ process after the full selection are 21 events and 2149 events, respectively.
The significance of $s/\sqrt{s+b}$, where s is the number of signal events ($\ttbb$) and b is the sum of $\ttjj$ and $\ttH$ background events, 
is 5.8 without taking into account the systematic uncertainty.

\section{DISCUSSION}

Differential kinematic distributions of the additional jets from the $\ttbb$ process 
are compared with the additional jets from the $\ttjj$ process and
with the two b jets from the Higgs boson from $\ttH$ events where $H$ decays to $b\bar{b}$ at $\sqrt{s}$ = 13 TeV.
In the best scenario where the b jet assignment is correct, 
the study clearly shows that the invariant mass and the $\Delta R$ variables of the two additional jets are the promising variables
for separating out those three processes.
In reality, identifying the origin of the b jets at the reconstruction level with 
the real data events remains challenging. 
Because in a real data analysis, 
no single best variable would be able to separate the $\ttH$ process from the $\ttbb$ or $\ttjj$ process,
this would require us to perform a multivariate analysis to increase the significance.
In the multivariate analysis, the variables that we presented here can be used as input variables.

We also discuss the significance of $\ttbb$ with the data corresponding to an integrated luminosity of 10 fb$^{-1}$ 
which is foreseen to be collected during the early Run II period.
The significances are 2.9 and 5.8 for the dileptonic decay mode, 
and the semileptonic decay mode, respectively.
However, if we take into account the systematic uncertainty of the $\ttbar$ cross section from Run 1 measurements,
5\% for the dileptonic events~\cite{dilepton} and
7\% for the semileptonic events~\cite{singlelepton},
the significance would go down.
From the following formula for the significance taking into account the systematic uncertainty,
\begin{equation}
s/\sqrt{s+b+(b \times unc.)^2},
\end{equation}
where $unc.$ indicates the systematic uncertainty of the background,
the significances go down to 1.9 for the dileptonic decay mode and 1.8 for the semileptonic mode.
The significances become compatible with each other for both channels.
This requires us to use a more complex approach 
and the precise measurements of $\ttbar$ pair production 
in order to maintain a high significance at $\sqrt{s}$ =  13 TeV with the data corresponding to an integrated luminosity of 10 fb$^{-1}$.  

\section{CONCLUSION}

We present a study of top-quark pair production in association with a bottom-quark pair
from fast simulations using the CMS detector.
The invariant mass and $\Delta R$ variable of two additional jets are promising variables
to separate $\ttbb$ events from $\ttjj$ and $\ttH$ events.
However, identifying the origin of the b jets at the reconstruction level with real data events
is challenging.
With the 10 fb$^{-1}$ data, in the best scenario with a simple cut and count method, 
we can reach 2.9 and 5.8 standard deviations for the dileptonic decay mode and the semileptonic decay mode, respectively. 
When the systematic uncertainty for the cross section measurement of the $\ttbar$ production is 
taken into account, 
the significance goes down to 1.9 for the dileptonic mode and 1.8 for the semileptonic mode, so 
the values of the significance becomes compatible with each other.
This would require us to measure the $\ttbar$ pair production cross section precisely at $\sqrt{s}$ = 13 TeV and to use a
more complex approach, such as a multivariate technique, for the $\ttbb$ cross section measurement.


\section*{ACKNOWLEDGMENTS}

This paper was supported by research funds of Chonbuk National University in 2014. 
This research was supported by the Basic Science Research Program through the National Research Foundation of Korea(NRF) funded by the Ministry of Education, Science and Technology (MEST, 2014R1A1A2056283).
This work was supported by a National Research Foundation of Korea Grant funded by the Korean governments (MEST, 2014S1A2A2028546).
This research was supported by the Basic Science Research Program through the National Research Foundation of Korea (NRF) funded by the Ministry of Science, 
ICT and Future Planning (2011-0016554) and funded by the Ministry of Education (NRF-2013R1A1A2062722). Also, it was supported by a Korea University grant.


\end{document}